\documentclass[aps,prb,twocolumn,showpacs,superscriptaddress,reprint]{revtex4-1}
\usepackage{graphicx}
\usepackage{xcolor}   
\usepackage{amsmath}
\usepackage{epsfig}

\begin{document} 
\title{Phonon transport in a one-dimensional harmonic chain with long-range interaction and mass disorder}
\author{Hangbo Zhou}
\affiliation{Institute of High Performance Computing, A*STAR, 138632, Singapore}
\author{Gang Zhang}
\email[]{zhangg@ihpc.a-star.edu.sg}
\affiliation{Institute of High Performance Computing, A*STAR, 138632, Singapore}
\author{Jian-Sheng Wang}
\affiliation{Department of Physics, National University of Singapore, 117551, Singapore}
\author{Yong-Wei Zhang}
\affiliation{Institute of High Performance Computing, A*STAR, 138632, Singapore}
\date{\today}

\begin{abstract}
Atomic mass and interatomic interaction are the two key quantities that significantly affect the heat conduction carried by  phonons. Here, we study the effects of long-range (LR) interatomic interaction and mass-disorder on the phonon transport in a one-dimensional harmonic chain with up to $10^5$ atoms. We find that while LR interaction reduces the transmission of low frequency phonons, it enhances the transmission of high frequency phonons by suppressing the localization effects caused by mass disorder.  Therefore, long-range interaction is able to boost heat conductance in the high temperature regime or in the large size regime, where the high frequency modes are important. 

\end{abstract}
\maketitle

\section{Introduction}
Due to the fast-emerging of low-dimensional materials, heat transport in nanostructures has attracted intensive interest in recent years. In bulk materials, the heat conduction is a diffusive process governed by Fourier's law. On the other side, in nanostructures, their size scales can be comparable to the mean free path or the coherence length of the phonons. Hence the heat transport is no longer simply diffusive \cite{Zhang2005}. In fact, it is affected by a mixture of many different factors such as random disorders, interaction ranges, and interfaces. As a result, phonon transport at the nanoscale becomes a complicated phenomenon. Since nanostructure provides basic building block for constructing phononic device \cite{Li2012} and phonon-based quantum computing \cite{Zhu2006, Harty2014}, a clear understanding of the effects of these various factors on phonon transport is highly demanded.

It is known that the atomic details of nanostructures are important in understanding the heat transport carried by phonons. Explicitly, atomic mass and interatomic interactions are the two major factors affecting phonon transport. In terms of atomic mass, phonons can be easily scattered by impurities or mass-disorders. In practice, mass-disordered materials are common, due to either unavoidable impurity contamination or intentional insertion of different atoms \cite{Nika2012,Sadeghi2012, Zhang2013, Lu2014}. The effects of mass disorder on thermal transport have been studied intensively using one-dimensional atomic chain models recently \cite{Zhao2006,Gaul2007,Dhar2008,Roy2008,Ni2011,liu2012,Ong2014,Ong2014a,Wang2015}. It is commonly believed that mass disorder causes localization of high-frequency modes, but it only weakly affects the transmission of low-frequency modes. In one-dimensional atomic chain with free boundary condition, the smallest localized frequency $\omega_c$ is size-dependent as $\omega_c\propto N^{-0.5}$, where $N$ is the number of atoms in the chain \cite{Ong2014a}. However, in terms of interatomic interactions, most of previous studies were based on models taking into account only nearest-neighbour (NN) interaction. In many situations, strong long-range (LR) interaction is present and able to affect phonon transport. For example, the ionic doping, such as  lithium insertion in nano structures for battery applications,  can produce long-range electrostatic potentials \cite{Xu2015, Li2015}. Strong long-range electric field can also be generated on fluorine-terminated surfaces as dipole lattices \cite{Mayrhofer2016}. Recently, in the design of quantum computer, long-range interacting phonons based on trapped ions are used  to create high-fidelity qubits for information transport \cite{Zhu2006, Harty2014}. Currently, the effects of LR interaction on phonon transport is not well understood. In the classical regime, it has been shown that even a next NN interaction can change the size-dependence of thermal conductivity when the phonons are scattered by anharmonicity \cite{Xiong2012}. However, when the phonons are scattered by mass disorder, the LR effects on quantum thermal conductance remain unexplored.

In this work, we employ a model of harmonic one-dimensional atomic chain with random masses to study the effects of long-range interaction on mass-disordered heat transport, as shown in Sec.~\ref{sec:model}. Such model is the simplest one that is able to capture both LR interaction and mass disorder. The more intricate scatterings due to anharmonicity are ignored and thus we focus on the effects of LR interaction on the phonons scattered only by mass disorder. Using the non-equilibrium Green's function (NEGF) technique, we reveal the frequency dependence of phonon transmission and also the size and temperature dependence of thermal conductance in Sec.~\ref{sec:results}. For systems without mass disorder, we found that LR interaction is able to reduce the low-frequency transmission, and hence reduce the thermal conductance. In contrast, for mass-disordered system in the presence of  the LR interaction, it is found surprisingly that LR interaction is able to release the localized phonon modes arising from mass-disorder, and hence increase the thermal conductance. Lastly, we studied different types and strengths of LR interaction, and analyzed their effects on thermal conductance.   

\section{Model and method}
\label{sec:model}

\begin{figure}
\includegraphics[width=\linewidth]{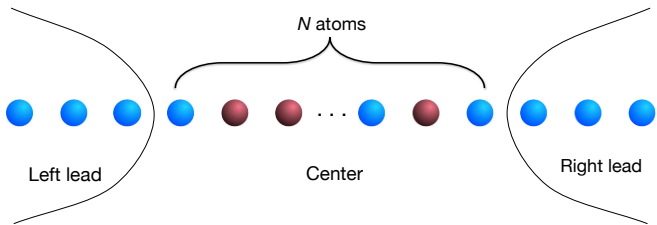}
\caption{\label{fig:model}A schematic of a one-dimensional mass-disordered system. The left and right leads are in equilibrium with different temperatures to drive a heat current. The atoms in the center have random mass disorder, labelled as red (dark gray). The atoms in the center have long-range interactions.}
\end{figure} 

We study heat transport through a system of a one-dimension atomic chain connecting to two phonon leads, as illustrated in Fig.~\ref{fig:model}. The leads are served as heat baths, providing temperature bias on the system. The total Hamiltonian can be written as:
\begin{equation}
H_{\mathrm{tot}}=H_L+H_{LC}+H_C+H_{CR}+H_R,
\end{equation}   
where $H_L$, $H_R$ and $H_C$ are the Hamiltonians of the left lead, right lead and central region. $H_{LC}$ and $H_{CR}$ are the coupling Hamiltonians between the central region and each leads, respectively. The Hamiltonian of the central region is given by:
\begin{equation}
H_C= \sum_{s=1}^N\frac{p_s^2}{2m_s}+V^C(x_1,x_2,\cdots,x_N),
\end{equation}
where index $s$ labels the atom in the center and $N$ is the total number of atoms, characterizing the size of the system. For atom $s$, $m_s$ is its mass, $x_s$ is its displacement operator and $p_s$ is its momentum. The first term denotes the kinetic energy contribution. The second term represents the potential energy due to interatomic interactions. Therefore the potential energy depends on the entire set of atomic displacements. Here we ignore the anharmonic effects so that the potential between each pair of atoms, $s$ and $s'$, are quadratic $\frac{1}{2}k_{ss'}(x_s-x_{s'})^2$, where the spring constant $k_{ss'}$ depends only on their relative distance. As a result, the total potential can be written as:

\begin{equation}
V^C=\sum_{1\le s<s'\le N}\frac{1}{2}k_{ss'}(x_s-x_{s'})^2+\frac{1}{2}k'(x_1^2+x_N^2),
\end{equation}
where the atoms at the two ends ($x_1$ and $x_N$) are connected to the heat baths. We assume that both the mass disorder and LR interaction only occur at the center so that the interaction between the system and leads, and the interaction within each lead are short-ranged so that only nearest neighbor couplings are taken into account. Here $k'$ is the spring constant coupled to the bath and it is constrained by the condition that the total Hamiltonian satisfies translational invariance. By rearranging the terms, the potential energy can be written in the matrix form as $V^C=\frac{1}{2}x^TK^Cx$, where $x$ is the column vector of displacement operators and $K^C$ is the force constant matrix.

Using the NEGF framework, heat conductance at temperature $T$ can be evaluated from the Landauer formula
\begin{equation}
\sigma=\int_0^\infty\frac{d\omega}{2\pi}\hbar\omega \mathcal{T}[\omega]\frac{\partial f}{\partial T},
\end{equation}
where $f=[\mathrm{exp}(\frac{\hbar\omega}{k_BT})-1]^{-1}$ is the Bose-Einstein distribution and $\mathcal{T}[\omega]$ is the transmission coefficients. It can be evaluated from the Caroli formula $\mathcal{T}[\omega]=\mathrm{Tr}[G^r(\omega)\Gamma_L(\omega)G^a(\omega)\Gamma_R(\omega)]$. $G^r$($G^a$) is the retarded (advanced) Green's function of the central region in the frequency domain \cite{Ni2011,Wang2008review},
\begin{equation}
G^r[\omega]=[(\omega+i\eta)^2M-K^C-\Sigma_L^r(\omega)-\Sigma_R^r(\omega)]^{-1}
\end{equation}
and $G^a(\omega)=[G^r(\omega)]^\dagger$. Here $M$ is the diagonal mass matrix, $\eta$ is a positive small number ($\eta\to 0^+$) and $\Sigma_\alpha^r(\omega)$ is the regarded self-energy of lead $\alpha$. The spectral function of the lead $\Gamma_\alpha$ is given by $\Gamma_\alpha(\omega)=-2\mathrm{Im}[\Sigma_\alpha^r(\omega)]$. In this work, we choose our baths as standard Rubin baths of force constant $k$ so that the regarded self-energy of each lead is given by  \cite{Wang2014review}
\begin{equation}
\Sigma^r_\alpha(\omega)=k'^2[\Omega+k-k'+k\lambda]^{-1},
\end{equation}
where $\Omega=m(\omega+i\eta)^2-2k$ and $\lambda=(-\Omega\pm\sqrt{\Omega^2-4k^2})/(2k)$ are two shorthand notations. The choice of the plus or minus sign is determined by the condition $|\lambda|<1$.

In order to generate mass disorder in the central region, we set each atomic site with a probability $c$ to have an impurity atom with mass $m'$. Consequently, a probability of $1-c$ is set for each atomic site with mass $m$. The atoms with mass $m$ are the same as those in the leads. When the number of atoms $N$ or the amount of ensembles is large, the concentration of impurity atoms converges to $c$. For each $N$, we average over 100 ensembles of random configurations during the calculation of the transmission coefficients, in order to achieve satisfactory convergent results.
  
In system with long-range interaction, the spring constant $k_{ss'}$ decays with respect to the relative distance $r_{ss'}$ between the atoms $s$ and $s'$. In this work, we cut off the interaction range until it is sufficiently small (less than 1\% of NN interaction strength). Here we consider two forms of long-range interaction, the inverse-square decay $k_{ss'}=k_0(r_{ss'}/a)^{-2}$ in analogue to Coulomb interaction; and exponential decay $k_{ss'}=k_0 d^{(1-r_{ss'}/a)}$ for the control of long-range interaction strength. Here $k_0$ is the NN spring constant, $a$ is the lattice constant and $d$ is the decay length to control the long-range interaction strength. The atomic distance is $r_{ss'}=|s-s'|a$.  For simplicity, we set $k'=k_0=k$ so that if one only considers the NN interaction without mass disorder, the system should restore to an ideal atomic chain, in which all the modes can transmit perfectly.  

\section{Results and discussion}
\label{sec:results}
\subsection{LR effects in homogeneous lattice}
\label{sec:nodisorder}   

\begin{figure}
\includegraphics[width=\linewidth]{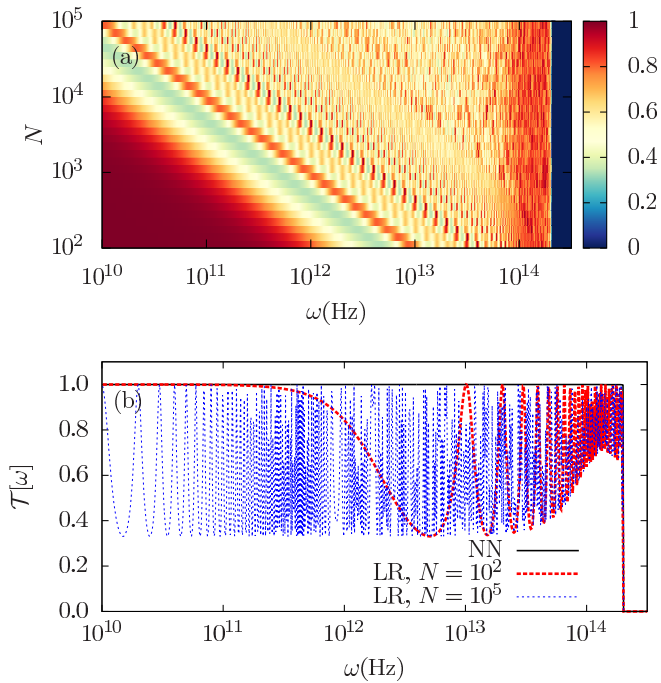}
\caption{\label{fig:nodisorder1}Transmission coefficients in a one-dimensional homogeneous lattice with long-range interaction under inverse-square decay. (a) The length and frequency dependence of transmission coefficients. (b) The frequency dependence of transmission coefficients for system with nearest-neighbor (NN) coupling (solid line) and  system with long-range (LR) coupling (dotted line). The parameters of spring constant $k_0$ and mass $m$ are chosen such that $\omega_0=\sqrt{k_0/m}=10^{14}$Hz and this applies to all the figures below.}
\end{figure}

We first investigate the properties of phonon transport when there is no mass disorder in the central region, by setting $c=0$. In case of considering only NN coupling, the transmission coefficient is equal to one for all the available modes, independent of the number of atoms $N$ in the central region. 

Next, we take the LR interaction into account. First, we consider the scenario that the LR interaction with interaction strength decays  inverse-squarely with respect to the interatomic distance. The transmission function, which is shown in Fig.~\ref{fig:nodisorder1}, presents significantly different behaviors in comparison with NN coupling. It starts to oscillate with respect to frequency and thus the transmission gets suppressed on average. 
The oscillation starts from high frequency since in the low frequency limit $\omega\to 0$, the transmission should in principle converge to 1 due to the translational invariance. The starting frequency of oscillation decreases with increasing size, proportional to $N^{-0.5}$. For larger $N$, the oscillation becomes denser, due to the increasing number of resonance modes of the system. At the high frequency regime, the cut-off frequency for the transmission function is invariant under LR interactions. 

In order to understand the physical mechanisms of the oscillations, we evaluated the oscillation peaks when the system is weakly coupled to the two leads, and found that they coincide with the sharp peaks at the resonance frequencies of the system. When the coupling becomes stronger, the peaks start to broaden and shift, affected by the self-energies of the leads. Hence, we conclude that such oscillations are due to the interference between the resonance modes of the system and the modes of the leads. The eigenmodes of the system will filter and select the incoming modes from the leads, affecting the transmission function.
\begin{figure}
\includegraphics[width=\linewidth]{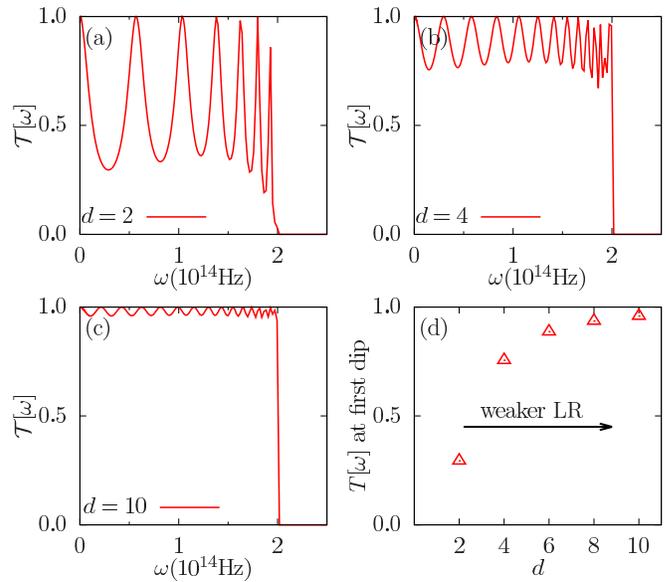}
\caption{\label{fig:nodisorder2}Plots of transmission coefficients in a one-dimensional homogeneous lattice with long-range interaction under exponential decay. The transmission coeffcient is plotted against frequency for (a) strong, (b) moderate and (c) weak long-range interaction strength. (d) The value of the transmission coefficients at the first dip counting from the left-hand side is plotted against the long-range interaction strength $d$. In this plot, the size is set to $N=20$.}
\end{figure}

To demonstrate that this oscillation of transmission function is independent on the format of LR coupling, we also study the harmonic chain in which LR interaction strength decays exponentially with respect to interatomic distance $k_{ss'}=k_0 d^{(1-r_{ss'}/a)}$. Here we keep $k_0$ as a constant but vary $d$ for all the calculations. A larger $d$ means that  the spring constant decays faster to zero, representing a weaker LR interaction, and it will restore to NN coupling when $d\rightarrow +\infty$. The oscillation in transmission coefficient is also observed in systems with this type of LR interaction. Furthermore, we investigate the relationship between such oscillations and the strength of long-range interaction by controlling the decay factor $d$ in the exponential form. The results are shown in Fig.~\ref{fig:nodisorder2}. We find that the LR interaction strength can significantly affect the oscillation amplitude. A stronger LR interaction strength has a greater capability to suppress the transmission coefficients. In order to give a clear picture of their relationship, we plot the transmission coefficient at the first dip of the oscillation against $d$. We find that the transmission coefficient increases exponentially with increasing $d$ and converges to 1. Hence, the size $N$ controls the starting frequency and the density of oscillation, while the LR interaction strength controls the amplitude of the oscillations. We also notice that both $N$ and $d$ do not affect the cut-off frequency, which is located around $\omega=2\times 10^{14}$Hz. This scenario will change if we introduce mass disorder, as illustrated in Sec.~\ref{sec:LRdisorder}. 

\begin{figure}
\includegraphics[width=\linewidth]{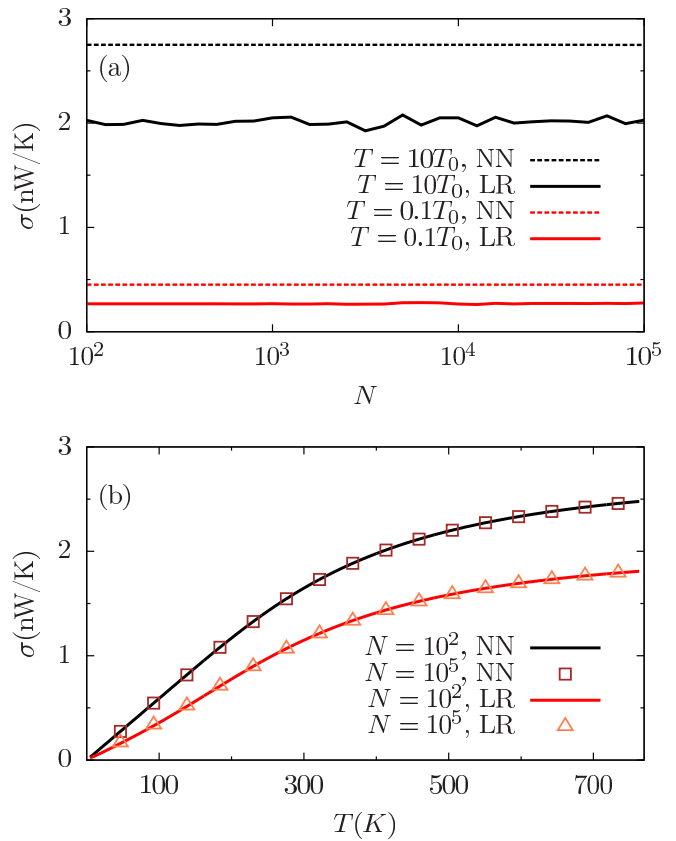}
\caption{\label{fig:nodisorderSigma}Thermal conductance for both long-range (LR) interacting system with inverse-square decay of interatomic interaction and nearest-neighbor (NN) interacting system. (a) The length dependence of thermal conductance at high temperature $T=10T_0$ (black or upper lines) and  low temperature $T=0.1T_0$ (red or lower lines), where $T_0=\hbar\omega_0/k_B\approx 764$K. (b) The temperature dependence of thermal conductance for short length $N=10^2$ and long length $N=10^5$. The form of long-range interaction here is inverse-square decay.}
\end{figure}

We then study the effects of LR interaction on thermal conductance. We first look at its size dependence as shown in Fig.~\ref{fig:nodisorderSigma}(a). In case of NN coupling, the thermal conductance does not depend on $N$ as the transmission coefficient itself is $N$ independent. Interestingly, in the case of LR coupling (inverse-square decay), though the weights of frequencies in transmission function depend heavily on $N$, the thermal conductance does not change significantly by changing the length,  in the tolerance of some fluctuations in the high temperature regime. In order to understand such behavior better, we investigate its temperature dependence since the effect of temperature is to scale the weights of different modes to the thermal conductance. At low temperature, low frequency modes contribute more while at high temperature, all modes contribute equally. However, from Fig.~\ref{fig:nodisorderSigma}(b), we still find that such length-independent behavior occurs at all temperature regime. This phenomenon is mathematically not obvious, but it can be physically explained. Since the increase in length does not introduce scattering process to the phonons, all the phonons are only scattered at the interfaces between the system and leads so that the thermal conductance should be length-independent. Such behaviors are expected to disappear if there exist either anharmonic scattering or mass-disorder scattering. However, it is interesting to find that with LR interaction, though the system size is not able to alter the overall thermal conductance, it does alter the transmission of individual modes. 

We also find that the thermal conductance of 1D harmonic chain with NN coupling is always larger than that with LR coupling for all ranges of atomic number $N$ and temperature $T$. However, this is only true when there is no mass disorder so that the NN coupling system has perfect transmission. When the system is mass-disordered, the system with LR interaction is more robust against the reduction of transmission coefficient, resulting in a different behavior in the thermal conductance. We elaborate this in the Sec.\ref{sec:LRdisorder}.  
\subsection{LR effects in system with mass disorder}  
\label{sec:LRdisorder}
\begin{figure}
\includegraphics[width=\linewidth]{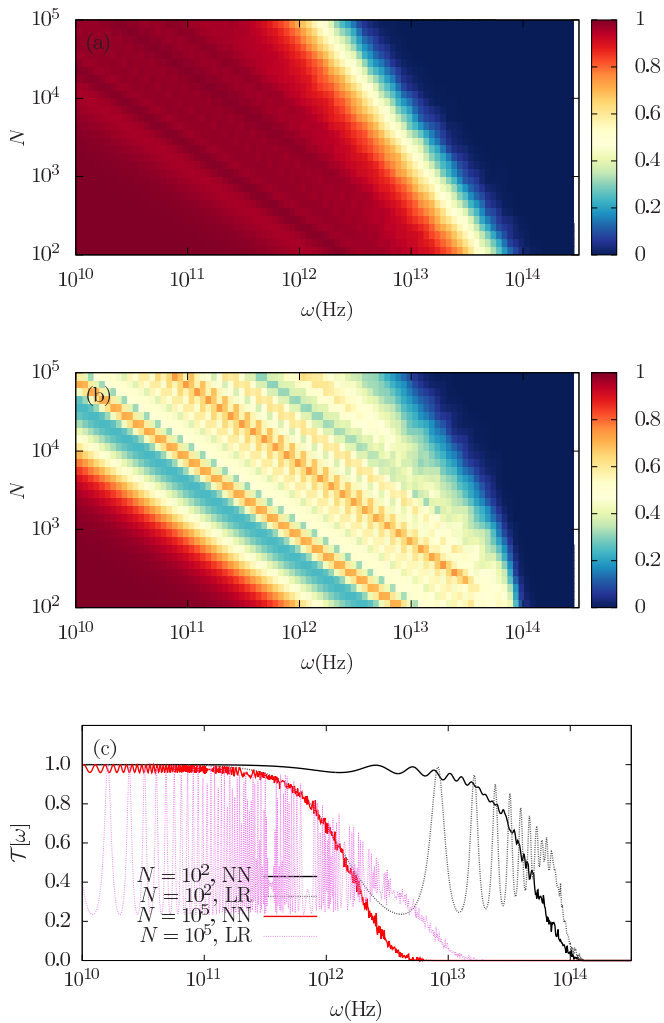}  
\caption{\label{fig:disorderT} Transmission coefficients for a system with mass disorder. (a) System only with nearest-neighbor coupling; and (b) system including long-range interaction with inverse-square decay. The impurity mass is $m'=2m$ with concentration $c=0.5$. (c) Plot of transmission coefficients against frequency.} 
\end{figure}

We next study the effects of LR coupling in a mass-disordered harmonic chain. The system is randomly doped with atoms of mass $m'=2m$, with a concentration of $c=0.5$. The transmission coefficient of such disordered system is shown in Fig.~\ref{fig:disorderT} for both NN coupling [panel(a)] and LR coupling [panel (b)].  In case of NN coupling, the presence of mass disorder makes the cut-off frequency $\omega_c$ red-shift and it further decreases by increasing the length $N$. Such phenomenon has been previously predicted \cite{Rubin1968, Matsuda1970,Ong2014,Ong2014a} and it is known that the cut-off frequency and the system size satisfy the relation $\omega_c\propto N^{-0.5}$. This effect is originated from the fact that the high frequency modes can be localized due to the mass disorder \cite{ Matsuda1970,Zhang2010}. However, once we introduce LR interaction with inverse-square decay [Fig.~\ref{fig:disorderT}(b)], the $N$ dependence of cut-off frequency changes with $N$. If we define $\beta$ as $\omega_c\propto N^{-\beta}$, we then find that $\beta>0.5$ when $N<10^{2.5}$ while $\beta\approx0.5$ when $N>10^{2.5}$. As a result, the LR coupling increases the cut-off frequency especially when $N$ is large, which can be clearly observed from Fig.~\ref{fig:disorderT}(c) in case of $N=10^5$ (red lines). As  discussed in Sec.~\ref{sec:nodisorder}, the increase in cut-off frequency is not observed when there is no mass disorder. This indicates that the LR coupling can boost the transmission of high frequency modes by releasing the localized modes arising from phonon scattering by mass disorder. Physically this is understandable because the localization induced by mass disorder is a short-range phenomenon, the range of which is only one lattice spacing. Since the long-range interaction allows each atom to interact with more atoms, it smears out the disorder. Thus the long-range interaction allows the local modes to interact with atoms with a longer distance, opening more channels for phonons to escape from localization and hence enhancing the transmission possibility.

It is known that the mass disorder is able to localize high frequency modes, but its effect on low frequency modes is insignificant. As discussed previously, LR coupling can suppress the low frequency transmission through oscillations. This phenomenon still exists in a mass-disordered system, as shown in Fig.~\ref{fig:disorderT}(c). These oscillations cannot be removed by taking ensemble average over different configurations of mass-disorders, indicating that it is intrinsic to the LR coupling. In summary, the effect of LR coupling is to reduce the transmission of low frequency modes, but enhance the transmission of high frequency modes by suppressing the mode localization. These effects will have important consequences on thermal conductance. 

\begin{figure}
	\includegraphics[width=\linewidth]{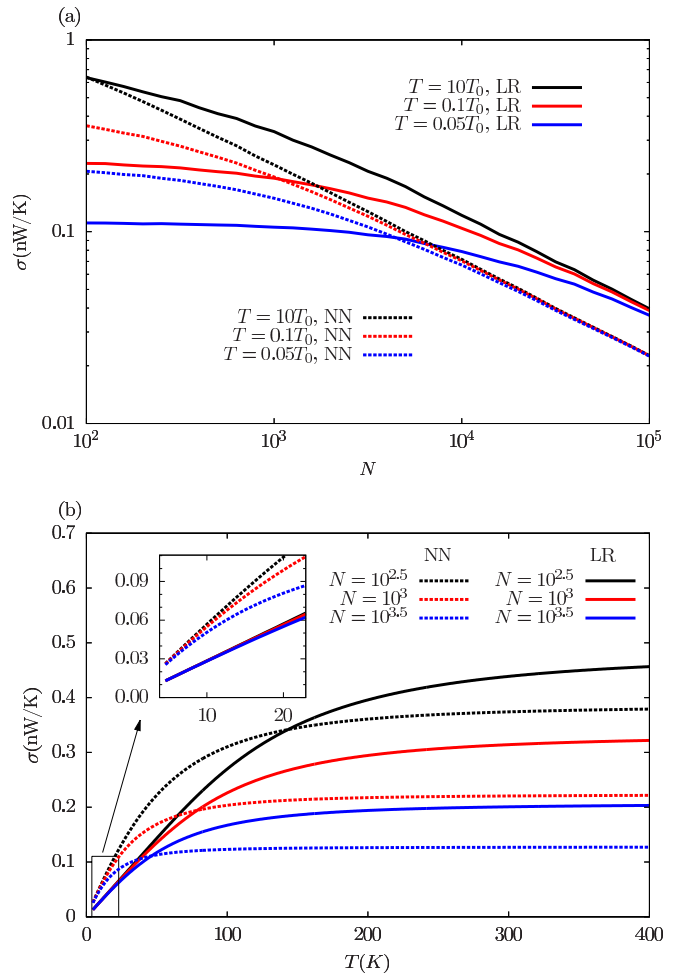}
	\caption{\label{fig:disorderSigma} (a) Length-dependence of thermal conductance at high (black or upper lines), moderate (red or middle lines) and low (blue or lower lines) temperatures. (b) The temperature dependence of thermal conductance at length of $N=10^{2.5}$ (black or upper lines), $N=10^3$ (red or middle lines) and $10^{3.5}$ (blue or lower lines).}
\end{figure}

Figure~\ref{fig:disorderSigma}(a) shows the length dependence of thermal conductance at different temperatures. For a fixed temperature, $T=0.05T_0$ or $T=0.1T_0$, we find that in the regime of short length, the thermal conductance of NN coupling is larger than that of LR coupling. However, with an  increase of length $N$, the thermal conductance of system consisting of LR coupling will eventually surpass those of NN coupling. This is because in the short length regime, the scatterings at the boundaries between the central region and the leads dominate over the mode localization caused by mass disorder. The NN coupling system has perfect transmission over the boundaries so it has a larger thermal conductance. However, for large $N$, the mode localization becomes significant. The LR interaction is able to suppress the localization effects and thus the thermal conductance becomes larger. Since only the high frequency modes are localized, such LR effect on the thermal conductance is more significant at higher temperature. This agrees with the fact that the crossover length decreases with increasing temperature. If we focus on the large $N$ regime, the thermal conductance becomes temperature-independent. For the NN coupling system, the thermal conductance decays as $\sigma\propto N^{-0.5}$ for all temperatures. This result is consistent with the previous predictions \cite{Ni2011, Dhar2001, Ong2014a} when anharmonic scattering is not taken into consideration. For a fixed $N$, the absolute value of the length dependence exponent with LR coupling becomes less than $0.5$. This is because LR coupling causes the cut-off frequency to respond reluctantly to length. However, the behavior at infinitely large $N$ limit is not predictable from the current results.

Fig.~\ref{fig:disorderSigma}(b) shows the temperature dependence of thermal conductance. For a fixed $N$, we find that there is a crossover temperature at which the thermal conductance of LR coupling system surpasses that of NN coupling system. This again agrees with the previous observation that the effect of LR coupling is to reduce the low frequency transmission while enhance the high frequency transmission. The crossover temperature decreases with the increase of $N$. We also find that when $N$ is larger, the thermal conductance saturates at a lower temperature. This is because a larger $N$ will make the cut-off frequency $\omega_c$ smaller. As a result, a low temperature is enough to excite all the available modes in order to make the heat conductance saturate. In the large $N$ limit, the saturation temperature tends to zero. This explains the observation in Fig.~\ref{fig:disorderSigma}(a) that the thermal conductance becomes independent of temperature $T$.

In the low temperature limit, we find that the thermal conductance is independent of $N$, which agrees with the fact that $N$ only affects the high frequency transmission. In this region, we also find that the thermal conductance increases linearly with $T$, $\sigma\propto T$, for both cases. However, the LR coupling can reduce the coefficient of linearity, which is caused by the reduction of transmission for low frequency modes. 

\subsection{Effects of LR interaction strength}
\label{sec:LRstrength}
 
 \begin{figure}
 	\includegraphics[width=\linewidth]{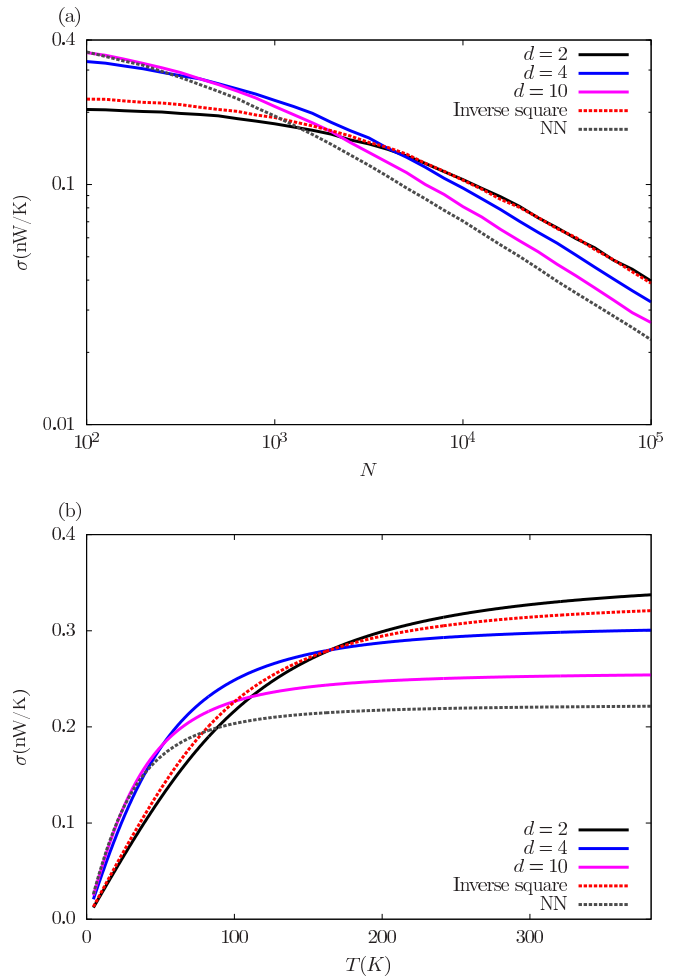}
 	\caption{\label{fig:decay} (a) The length-dependence of thermal conductance at different LR interaction strengths. The temperature is set at $T=0.1T_0$. (b) The temperature dependence of thermal conductance at different LR interaction strengths. The length is set at $N=10^3$.}
 \end{figure}
 
In this section, we study the effects of different forms and strengths of LR interaction. We focus on two forms of long-range interaction: exponential decay and inverse-square decay, as presented in Sec.~\ref{sec:model}. In Fig.~\ref{fig:decay}, we plot the thermal conductance under various LR interaction strengths $d=2, 4, $ and 10. It is noted that a larger $d$ means weaker LR interaction. When plotting with respect to length $N$ [Fig.~\ref{fig:decay}(a)], we find that system with a strong LR interaction has a smaller thermal conductance in the short length regime. However, when the length is long enough ($N>10^4$), the thermal conductance of strong LR interaction system will eventually surpass that of weak LR interaction system. This is again due to the competition between boundary scattering and impurity scattering. This indicates that a system with a stronger LR interaction has larger boundary scatterings as well as stronger capability of releasing the localized modes. When the system size is longer, $\mathrm{log}\sigma$ gradually becomes linearly dependent on $\mathrm{log}N$. We find that it is more difficult for systems with stronger LR interaction to approach this asymptotic behavior, indicating that the boundary effects are more challenging to overcome in a stronger LR interaction system.

Figure.~\ref{fig:decay}(b) shows the plot of thermal conductance with respect to temperature. We find that system with a stronger LR interaction has a higher thermal conductance at a high temperature but a lower thermal conductance at a low temperature. This result further supports the previous conclusion that a strong LR has a large capability to release the localized modes. At a high temperature, the contribution of high frequency modes becomes important. Thus the relaxation of localized modes has more significant influence on thermal conductance. As a result, in high temperature regime, system with a strong LR interaction has a larger thermal conductance. On the other hand, at a low temperature, low frequency modes contribute more to thermal conductance. This result is consistent with the previous prediction that LR interaction is able to suppress the transmission of low frequency modes, resulting in a reduced thermal conductance in low temperature regime.

Another important observation from Fig.~\ref{fig:decay} is that the thermal conductance shows a similar behavior for both inverse-square decay and exponential decay. The curve of inverse-square decay is sandwiched between the exponential decay curve of $d=2$ and $d=4$. Interestingly, these three curves share the same crossover point when plotting against temperature (at $T\approx 175$K). This phenomenon does not occur for any three exponential decay curves, indicating that the form of decay plays a role in the transition behavior but not the overall trend.

\section{Conclusion} 
We have investigated the effect of long-range interaction on the phonon transport in a one-dimensional harmonic chain with and without mass disorder. It is found that the long-range interaction is able to suppress the transmission of low-frequency modes arising from boundary scatterings. Meanwhile, it is able to enhance the transmission of high-frequency modes by suppressing the mode localizations arising from mass disorder. Consequently, long-range interaction is able to reduce the thermal conductance in a short length system while enhance it for a long length one. The crossover length increases with decreasing temperature. Similarly, system with long-range interaction is able to reduce the thermal conductance in the low temperature regime while enhance it in the high temperature regime. We have also studied the effects of different forms and strengths of long-range interaction on the behavior of phonon transport, and found that the drawn conclusion is robust under these different conditions. The presented temperature and size dependence of thermal conductance is experimental measurable by using trapped ion technique \cite{Smith2016}.

Our work reveals insights into the phonon transport behavior in a harmonic chain with LR interaction and mass disorder.  With such numerical study, it is possible to go beyond the disorder scatterings to investigate the LR interaction effects with anharmonic scatterings. The anharmonicity couples the phonon modes of different frequencies and it has been shown that the anharmonic scatterings introduced effective extended states and hence has great impacts on the localization in one-dimension classical chain \cite{Dhar2008b}. Furthermore, the anharmonicity will introduce another length scale, the mean free path of the phonons. Our work reveals the effect of LR interaction in the case when the mean free paths of the phonons are much longer than the interaction range. Hence future investigations on effects of long-range interaction when the mean free path approaches the interaction range are of great interest. In the classical regime, detailed investigation have been performed in anharmonic chains with LR interactions in order to change the heat current and improve the thermal rectification properties  \cite{Pereira2013,Avila2015, Chen2015}, implying that using LR interaction provides a promising routine for thermal managements.

\section*{Acknowledgment}
This work was supported in part by a grant from the Science and Engineering Research Council (152-70-00017). The authors gratefully acknowledge the financial support from the Agency for Science, Technology and Research (A*STAR), Singapore and the use of computing resources at the A*STAR Computational Resource Centre, Singapore.

\bibliography{LR}
\end{document}